\documentclass[12pt]{iopart}
\usepackage[utf8]{inputenc}
\begin{document}

\title[]{Quantum master equation approach to  heat transport  in dielectrics and semiconductors}
\author{Yamen Hamdouni}
\address{Physics department, Mentouri university, Constantine, Algeria}
\ead{hamdouniyamen@gmail.com}
\vspace{10pt}

\begin{abstract}
We report on the derivation of the heat transport equation for nonmetals  using a quantum Markovian master equation in Lindblad form. We first  establish the equations of motion describing the  time variation of the on-site energy of atoms in a one dimensional periodic chain that is coupled to a heat reservoir. In the  continuum limit, the Fourier law of heat conduction naturally emerges, and the heat conductivity is explicitly obtained. It is found that the effect  of the heat reservoir on the lattice  is described   by a heat source density  that depends on the diffusion coefficients of the atoms. We show that the Markovian dynamics is equivalent to the long wavelength approximation for phonons, which is typical for the case of elastic solids. The high temperature limit  is shown to reproduce the classical heat conduction equation.  
\end{abstract}

%
\vspace{2pc}
\noindent{\it Keywords}: open quantum systems,  Markovian dynamics, heat transport

\maketitle
\section{Introduction}
Thermal conductivity in solids has various origins depending on whether they are  metals or nonmetals~\cite{kittel,zim}. In the former case,  electrons are the main carriers of heat. In semiconductors and dielectrics, however,  phonons play the crucial role in the process of heat transport. The study of these properties has a significant importance especially in the use and the  fabrication of many devices that are based on these materials. The description of the dynamics of atoms in solids  usually relies on  their fundamental symmetries such as lattice periodicity, which has great consequences. Actually, phonons are merely the quantized perturbations that propagate throughout the periodic lattice. In their motion, phonons may undergo many interaction processes,  leading to their decay or relaxation. The latter fact affects significantly the phenomenon of heat transport by phonons. 

Generally speaking, heat conduction occurs when   a non-zero gradient of the temperature  exists in the material. This process is phenomenologically well described by the Fourier law which involves the so-called heat conductivity. The study of this property is essentially based on the Focker-Planck equation, where one is interested in the change of the distribution of the carriers in space and time~\cite{zim2,klem}. In either case, whether for electrons or phonons, the heat conductivity is derived using  the {\it relaxation time} ansatz, which yields for phonons a conductivity that is quite similar in its form to the conductivity of gases.

 In recent years, attention has been given to the quantum  description of heat transport in various physical systems~\cite{segal, dubi, chen, dhar, clark, asad, motz, olivera,novak,yan, zhou,hofer1}. Apart from the need for new theoretical paradigms and elaborated mathematical methods to handle the dynamics  in many-body systems in a purely quantum manner, this kind of problems has been associated with the development   of new experimental techniques that yield more precise and accurate measurements at the atomic and molecular scale  at low temperatures~\cite{pekola}. In addition, several works   have been  devoted to the study of quantum thermodynamics and  quantum thermal machines \cite{quan, abah, hofer2,leggio,hardal,hofer3,binder}, and quantum control~\cite{aless},  where the laws of quantum mechanics come to play. Besides, the question of how do classical laws of heat transport emerge from the quantum ones  is thus of fundamental importance.

 The main aim of this paper is to approach the problem of heat transport in nonmetals from the perspective of the theory of open quantum systems~\cite{petru}. 
 Indeed, given that we are dealing with a non-equilibrium problem~\cite{weiss}, we assume that the lattice is in contact with a heat reservoir and we propose to study the lattice dynamics using the quantum master equation technique~\cite{hamd1,hamd2}. The aim here is two-fold: First, we seek  to derive the equations of motion for the  constituents of the discrete lattice on the atomic level, and to draw conclusions regarding the way the energy decays and flows from one site to another. Second, we treat the classical heat transport as a limiting case of the discrete one by invoking the continuum limit, from which the heat conductivity of the lattice may be obtained in a nontrivial manner.
A natural choice for the continuum limit is the Markovian approximation for the master equation. The most general form of the latter has been found by Lindblad in his seminal work~\cite{lind1,lind2}. Practically, most of the Master equations used in various fields may be reduced to the Lindblad form. We shall use the latter by postulating a particular choice for the Lindblad operators that is best adapted to Harmonic motion of the atoms in the lattice.  It is worthwile to notice here that the use of the master equation formalism has been adopted for studying heat flow in spin chains by many authors. 

The paper is organized as follows. In Sec.~\ref{sec2}, we introduce the model and we derive the equations of motion for the on-site energy. Section~\ref{sec3} is devoted to the continuum limit, where we derive the classical heat transport equation, and deduce the heat conductivity. There, we discuss the implication of the Markovian approximation, and we study the high temperature limit. The paper is ended with a brief conclusion.  
\section{Model and equations of motion\label{sec2}}

In nonmetallic materials, heat transport is mainly due to the  out of equilibrium dynamics of phonons.  The contribution of electrons being negligible compared to that of the phonons, we disregard the electronic degrees of freedom of the atoms and, for the sake of simplicity, we restrict ourselves to one dimension. Under these conditions, we model the system of interest by the Hamiltonian:

\begin{equation}
 \hat H=\sum_{k=1}^N \frac{\hat p_k^2}{2m}+\left(\frac{1}{2} m\omega_0^2
+\xi\right) \hat x_k^2-\xi  \hat x_k \hat x_{k+1},\end{equation}
where all the atoms are assumed identical (monoatomic chain) with the same mass $m$. The frequency $\omega_0$ describes the local harmonic oscillator potential in which each atom is immersed, whereas the  parameter $\xi$,  which is essentially positive (i.e. $\xi>0$), represents the coupling constant between neighboring atoms. 
The operators $\hat x_k$ and $\hat p_k$ satisfy the usual canonical commutation relations:
\begin{equation}
 [\hat x_k,\hat p_{k^{'}}]=i\hbar \delta_{kk^{'}}, \qquad  [\hat x_k,\hat x_{k^{'}}]=0 , \qquad  [\hat p_k,\hat p_{k^{'}}]=0.
\end{equation}

In order to deal with the irreversible energy transport on the atomic scale, we assume that the system is weakly  coupled to  a heat reservoir, whose size and properties are not altered by the interaction with the lattice. This causes energy exchange between the chain and the reservoir as well as heat flow through the former until it reaches its equilibrium state. Furthermore, we assume that the time scale at which the state of the lattice changes is much longer than any time scale characterizing the reservoir. The latter conditions lead  to a Markovian-type  dynamics that is well described by means of a master equation in Lindblad form for the system's reduced density matrix $\hat \rho(t)$, which can be written in the Schr\"odinger picture  as~\cite{lind1,lind2}
\begin{equation}
\frac{d\hat\rho(t)}{dt}=-\frac{i}{\hbar}[\hat H,\hat\rho(t)]+ \frac{1}{2\hbar}\sum_\ell([\hat V_\ell \hat\rho(t),\hat V_\ell^\dag]+[\hat V_\ell, \hat\rho(t)\hat V_\ell^\dag])\label{mas1}.
\end{equation}
The influence of the heat reservoir on the state of the lattice is encoded in the Lindblad operators $\hat V_\ell$, which play a crucial role in the theory of open quantum systems. In the Heisenberg picture, the dual  evolution equation for the operator $\hat A$ reads: 
\begin{equation}
 \frac{d\hat A(t)}{dt}=\frac{i}{\hbar}[\hat H,\hat A(t)]+\frac{1}{2\hbar}\sum_\ell([\hat V_\ell^\dag[ \hat A(t),\hat V_\ell]+[\hat V_\ell^\dag, \hat A(t)]\hat V_\ell]).\label{mas2}
 \end{equation}

 We demand that the quantum  equations of motion in the presence of dissipation and damping be linear. This can be achieved by writing the Lindblad operators in the form of  linear combinations of the atoms coordinates and momenta operators, namely:
\begin{equation}
\hat V_\ell=\sum_{j=1}^N (a^\ell_j \hat p_j+b_j^\ell \hat x_j),\quad \hat V^\dag_\ell=\sum_{j=1}^N (a^{\ell*}_j \hat p_j+b_j^{\ell*} \hat x_j).
\end{equation}
The complex numbers  $ a^\ell_j$ and $b_j^\ell$ serve to introduce the quantum transport coefficients~\cite{hamd1}:
\begin{eqnarray}
D_{x_k x_j}&=&\frac{\hbar}{2}{\rm Re}\sum_\ell a^{\ell*}_k a^\ell_j, \quad  D_{p_kp_j}=\frac{\hbar}{2}{\rm Re}\sum_\ell b^{\ell*}_k b^\ell_j, \label{coef1}\\
D_{x_kp_j}&=&-\frac{\hbar}{2}{\rm Re}\sum_\ell a^{\ell*}_k b^\ell_j, \quad \lambda_{kj}=-{\rm Im}\sum_\ell a^{\ell*}_k b^\ell_j,\label{coef2}\\
\alpha_{kj}&=&-{\rm Im}\sum_\ell a^{\ell*}_k a^\ell_j,\qquad \eta_{kj}=-{\rm Im}\sum_\ell b^{\ell*}_k b^\ell_j,\label{coef3}
\end{eqnarray}
which enables us to rewrite the master equation~(\ref{mas2}) in the more explicit form:
\begin{eqnarray}
\fl \frac{d\hat A}{dt}=\frac{i}{\hbar}[\hat H,\hat A]+\frac{1}{2\hbar^2}\sum_{kj}\Biggl\{(i\hbar\alpha_{kj}-2D_{x_kx_j})\Bigl(\{\hat A,\hat p_k\hat p_j\}-2\hat p_k \hat A\hat p_j\Bigl)\nonumber\\+ (i\hbar\eta_{kj}-2D_{p_kp_j})\Bigl(\{\hat A,\hat x_k\hat x_j\}-2\hat x_k \hat A\hat x_j\Bigl)+(2D_{p_kx_j}+i\hbar\lambda_{jk})\times \nonumber\\\Bigl(\{\hat A,\{\hat p_j, \hat x_k\}\}-2(\hat p_j \hat A \hat x_k+\hat x_k \hat A \hat p_j)\Bigl)-2\hbar\lambda_{jk}\Bigl(\{\hat A,\hat p_j\hat x_k\}-2\hat x_k\hat A\hat p_j\Big)\Biggr\}.\label{master}
\end{eqnarray}
In the above equation, $\{\hat A,\hat B\}$ designates the anticommutator of the operators $\hat  A$ and $\hat B$; the corresponding  expectation values $\overline A$ and  $\overline {AB}$ are defined by:
\begin{eqnarray} 
\overline{A}(t)&=&{\rm tr}(\hat \rho \hat A(t)),\\ \overline {AB}(t)&=&\frac{1}{2}{\rm tr}\Bigl(\hat \rho\{\hat A(t),\hat B(t)\}\Bigl),
\end{eqnarray}   
where ${\rm tr} (.)$ denotes the trace with respect to the lattice state.

Using the master equation ({\ref{master}), it can be shown that the evolution in time of the expectation values of the bilinear terms involved in the lattice Hamiltonian satisfy:
\begin{eqnarray}
\fl \frac{d\overline{ x_k^2}}{dt}=-2\lambda_{kk}\overline {x_k^2}+\frac{2\overline {p_k x_k}}{m}-2\sum_{j\neq k}\lambda_{kj}\overline{x_k x_j}-2\sum_{j\neq k}\alpha_{kj}\overline {x_k p_j}+ 2D_{x_kx_k},\\
 \fl \frac{d\overline{ p_k^2}}{dt}=-2\lambda_{kk}\overline {p_k^2}-\left( m\omega_0^2
+2\xi\right) \overline {x_kp_k}-2\sum_{j\neq k}\lambda_{kj}\overline{p_k p_j}\nonumber \\ +2\sum_{j\neq k}[\eta_{kj}+\xi(\delta_{j k-1}+ \delta_{j k+1})] \overline {p_k x_j}+2D_{p_kp_k},\\
 \fl \frac{d\overline{ x_k x_{k+1}}}{dt}=-(\lambda_{kk}+\lambda_{ k+1 k+1})\overline {x_k x_{k+1}}+\frac{1}{m}(\overline {x_kp_{k+1}}+\overline {x_{k+1}p_k})\nonumber\\-\sum_{j\neq k}\lambda_{kj}\overline{x_{k+1} x_j}-\sum_{j\neq k+1}\lambda_{k+1j}\overline{x_{k} x_j}\nonumber \\-\sum_{j\neq k}\alpha_{kj}\overline {x_{k+1} p_j}-\sum_{j\neq k+1}\alpha_{k+1j}\overline {x_{k} p_j}+2D_{x_kx_{k+1}},
\end{eqnarray}
where we have assumed periodic boundary conditions $x_1=x_{N+1}$.

We  shall define the expectation value of   the on-site energy  by
\begin{equation}
 E_k= \frac{\overline {p_k^2}}{2m}+\left(\frac{1}{2} m\omega_0^2
+\xi\right) \overline {x_k^2}-\frac{\xi}{2} ( \overline {x_k x_{k+1}}+  \overline {x_k x_{k-1}}).
\end{equation}
The above quantity displays the non-local exchange nature of the energy, namely via the coupling terms   $ \overline {x_k x_{k\pm 1}}$, leading thus to energy transport through the lattice. 
Since  all the atoms are identical, we set $\lambda_{kk}=\lambda$ for all $k$, and  by the antisymmetry character  of the coefficients $\alpha_{kj}$ and $\eta_{kj}$, we have  $\alpha_{kj}=0$, $\eta_{kj}=0$. The friction coefficients $\lambda_{kj}$, on the other hand are symmetrical, and if we restrict the coupling to the nearest neighbors, we may write them in the form:
\begin{equation}
 \lambda_{kj}=\gamma(\delta_{jk+1}+ \delta_{jk-1}), \qquad {\rm for} \quad  j\neq k.
\end{equation}
Consequently,
\begin{eqnarray}
 \frac{dE_k}{dt}&=&-2\lambda E_k +\frac{D_{p_kp_k}}{m}+\left( m\omega_0^2
+2\xi\right) D_{x_kx_k}-\xi(D_{x_kx_{k+1}}+ D_{x_kx_{k-1}})\nonumber \\
&-&\frac{\xi}{2m}\left(\overline{x_k p_{k+1}}+ \overline{x_k p_{k-1}}-\overline{x_{k+1}p_k }-\overline{ x_{k-1}p_k}\right)\nonumber \\
&-&\frac{\gamma}{m}(\overline{p_k p_{k+1}}+\overline{p_k p_{k-1}})-\gamma(m\omega_0^2+2\xi)(\overline{x_k x_{k+1}}+\overline{x_k x_{k-1}})\nonumber \\
&+& \frac{\gamma \xi}{2} \left( 2 \overline{x_k^2}+\overline{x_{k+1}^2}+ \overline{x_{k-1}^2}\right)\label{image}.
\end{eqnarray} 
The right-hand side of equation (\ref{image}) may be regarded as describing  two-particle and  three-particle scattering processes, by which  the lattice atoms indirectly  exchange energy through the environmental particles. Moreover, it is worthwhile mentioning that the term
\begin{equation}
 \frac{\xi}{2m}\left(\overline{x_k p_{k+1}}+ \overline{x_k p_{k-1}}-\overline{x_{k+1}p_k }-\overline{ x_{k-1}p_k}\right)=a \nabla_a J_k \label{grad1}
\end{equation}
is the algebraic sum of   microscopic energy currents, which manifests itself in the form of a discrete gradient of a local  current $J_k$ on the lattice, where $a$ is the lattice constant:
\begin{eqnarray}
 a \nabla_a J_k=J_{k+1}-J_k,\\
 J_k=\frac{\xi}{2m}\left(\overline{x_{k-1} p_{k}}- \overline{x_{k} p_{k-1}}\right).
\end{eqnarray}
For the sake of brevity, it is convenient to rewrite equation~(\ref{image}) as:
\begin{eqnarray}
 \frac{dE_k}{dt}&=&-2\lambda E_k +\frac{D_{p_kp_k}}{m}+\left( m\omega_0^2
+2\xi\right) D_{x_kx_k}\nonumber \\&-&\xi(D_{x_kx_{k+1}}+ D_{x_kx_{k-1}})
+\mathcal L_k \label{old},
\end{eqnarray}
where
\begin{eqnarray}
\mathcal L_k&=&-\frac{\xi}{2m}\left(\overline{x_k p_{k+1}}+ \overline{x_k p_{k-1}}-\overline{x_{k+1}p_k }-\overline{ x_{k-1}p_k}\right)\nonumber \\
&-&\frac{\gamma}{m}(\overline{p_k p_{k+1}}+\overline{p_k p_{k-1}})-\gamma(m\omega_0^2+2\xi)(\overline{x_k x_{k+1}}+\overline{x_k x_{k-1}})\nonumber \\
&+& \frac{\gamma \xi}{2} \left( 2 \overline{x_k^2}+\overline{x_{k+1}^2}+ \overline{x_{k-1}^2}\right).
\end{eqnarray} 
The main contribution to $\mathcal L_k$ comes from the quantity displayed in the first line of the above equation, whose characteristic  time scale is of the order of:
\begin{equation}
 t\sim \tau=\frac{1}{2\lambda}.
\end{equation}
This shows that the effect of any perturbation at site $k$ may extend  beyond the lattice constant $a$ over which the gradient in equation~(\ref{grad1}) is defined. The range  actually depends on the speed of propagation of the perturbation. We shall return to this point later when we study the continuum limit. The contribution of the remaining terms of $\mathcal L_k$ is generally small because $\gamma\ll \lambda$; hence, they  can be neglected in the continuum limit (see bellow).

The lattice ultimately reaches its equilibrium state which  is a Gibbs state at the final temperature $T_f$. In this case the diffusion coefficients do not depend on $k$ and are the same for all the atoms; we express this fact symbolically by:
\begin{eqnarray}
 D_{x_kx_k}&=&D_{xx}(T_f),\qquad  D_{p_kp_k}=D_{pp}(T_f),\label{dif1}\\
  D_{x_kx_{k-1}}&=& D_{x_kx_{k+1}}=D_{\rm ex}(T_f).\label{dif2}
\end{eqnarray}

\section{Continuum limit\label{sec3}}
The passage to the continuum limit should take into account the elasticity  of the lattice and the way energy flows from point to point, namely the position dependence of the energy current. Hence when the lattice is out of equilibrium, the energy should be dealt with as a continuously varying quantity whose rate of change in time  is tightly related to its spacial variation, that is to its gradient. To this end let us introduce the quantity:

\begin{equation}
 u_k=\frac{E_{k}}{a}.
\end{equation}
 In the continuum limit, $u_k$ becomes the energy density (i.e. energy by unit of length) of the lattice, which will be denoted from here on by $u(x,t)$:
\begin{equation}
 \frac{E_k}{a}\Bigl|_{x=ka}\rightarrow  u(x,t).
\end{equation}

Let  $b$ denote the range  over which the  gradient of the energy changes in the homogeneous lattice  during a time interval of the order of the decay time constant:
\begin{equation}
 \tau =\frac{1}{2\lambda}.
\end{equation}
 Therefore, the variation of  the energy  current density at site $k$ reads:
 \begin{equation}
 \Delta\dot u_k= 2\lambda (u_{k+[\frac{b}{a}]}-u_k)-2\lambda (u_k-u_{k-[\frac{b}{a}]}),
 \end{equation}
  where $[x]$ designates the round value of $x$. The above  variation captures the change of the energy over the range $b$, including the contribution of the  nonlocal terms $\mathcal L_k$ in the continuum limit, where:
  \begin{equation}
  \Delta\dot u_k \rightarrow 2\lambda b^2 \frac{\partial^2 u(x,t)}{\partial x^2}.
  \end{equation}
  This is to be added to the decay of the energy during the time interval  $\tau$, which is described by the damping coefficient $2\lambda$. It follows that in the continuum limit equation (\ref{old}) turns into:
\begin{eqnarray}
 \frac{\partial u(x,t)}{\partial t}=2\lambda b^2 \frac{\partial^2 u(x,t)}{\partial x^2}-2\lambda u(x,t)+  s(T_f)\label{heateq1}
\end{eqnarray}
where 
\begin{equation}
   s(T_f)= \frac{1}{a}\left[ \frac{D_{pp}(T_f)}{m}+\left( m\omega_0^2
+2\xi\right)D_{xx}(T_f)-2\xi D_{\rm ex}(T_f)\right].\label{sink}
\end{equation}
In fact equation~(\ref{heateq1}) may be written as 
\begin{equation}
  \frac{\partial u(x,t)}{\partial t}= \frac{\partial }{\partial x} \left( 2\lambda b^2 \frac{\partial u(x,t)}{\partial x}\right) -2\lambda u(x,t)+   s(T_f)\label{heateq2}.
\end{equation}
This is the transient heat transport equation with source density $  s(T_f)$. The corresponding heat current reads:
\begin{equation}
 J(x,t)= -2\lambda b^2 \frac{\partial u(x,t)}{\partial x}.
\end{equation}
The latter can  actually be put in the more familiar form:
\begin{equation}
 J(x,t)=-\kappa\frac{\partial T}{\partial x}\label{fourier1},
\end{equation}
where the thermal conductivity is given by:
\begin{equation}
 \kappa=2\lambda b^2 C(T), 
\end{equation}
with 
\begin{equation}
 C(T)=\frac{\partial u(x,t)}{\partial T}
\end{equation}
 being the heat capacity density of the lattice.
\subsection{Range of propagation and the Markovian approximation} 
The derivation of the heat transport equation~(\ref{heateq2}) and the  thermal conductivity $\kappa$ was based on a Markovian master equation. The obtained results imply that the speed of propagation  of heat is constant. We may  thus be  tented to introduce the effective velocity 
\begin{equation}
 v_{\rm eff}=\frac{b}{\tau},
\end{equation}
and to write the thermal conductivity in the form:
\begin{equation}
 \kappa= v_{\rm eff} b C(T).
\end{equation}
The latter form is exactly the same as that corresponding to thermal conductivity in classical gases. This analogy, however, should be dealt with more care, because we are actually dealing  with phonon propagation. The study of this phenomenon so far has been based on a Focker-Planck equation together with the so-called relaxation time ansatz. The heat conductivity depends on the phonons modes labeled by the wave vectors $\mathbf q$, and can be written in one dimension as~\cite{klem}:
\begin{equation}
 \kappa =\frac{1}{N a}\sum_{\mathbf q} v(\mathbf q) r(\mathbf q)  \frac{\partial \epsilon(\mathbf q,T)}{\partial T},
\end{equation}
where  $\epsilon(T,\mathbf q)$ is the Bose-Einstein distribution,  $v(\mathbf q)$ is the group velocity of mode $\mathbf q$ and $r(\mathbf q)=v(\mathbf q)\tau (\mathbf q)$ is the range of that mode, with $\tau(\mathbf q)$  being the corresponding relaxation time.

Therefore, the Markovian approximation combined with the elastic continuum limit corresponds to  constant group velocity of the phonons. This happens in the long wavelength approximation for which only phonons of low frequencies do contribute to the dynamics. Consider for instance the simplified model where $\omega_0=0$. The modes frequencies are given by:
\begin{equation}
 \omega(\mathbf q)=2\sqrt{\frac{\xi}{m}}|\sin(qa/2)|.
\end{equation}
In the long wavelength approximation, associated with the continuum limit, we have $qa\ll 1$, meaning that:
\begin{equation}
 \omega(\mathbf q)=a\sqrt{\frac{\xi}{m}} q,
\end{equation}
which yields a constant group velocity for all involved modes:
\begin{equation}
 v(\mathbf q)=a\sqrt{\frac{\xi}{m}}.
\end{equation}
The latter is the velocity of sound in the lattice, and is the actual speed of propagation of heat throughout the lattice. The range $ b$ can thus be written in the Markovian approximation as
\begin{equation}
 b=a\sqrt{\frac{\xi}{m}} \tau=\frac{a}{2\lambda} \sqrt{\frac{\xi}{m}},
\end{equation}
yielding the heat conductivity
\begin{equation}
 \kappa=\frac{ a^2 \xi}{2\lambda m}C(T).
\end{equation}
More explicitly, we can write:
\begin{eqnarray}
 \frac{\partial u(x,t)}{\partial t}=\left(\frac{ a^2 \xi}{2\lambda m}\right) \frac{\partial^2 u(x,t)}{\partial x^2}-2\lambda u(x,t)+  s(T_f).
\end{eqnarray}
Hence we conclude that the continuum heat transport equation emerges from the combination of the long wavelength and the  Markovian approximations  when applied to the quantum master equation.

 \subsection{High temperature  limit}
 For simplicity we assume that $\gamma=0$; the case of $\gamma\ne 0$ is dealt with in the appendix, and the outcome is essentially the same. The temperature dependence of the source $s(T_f)$ for arbitrary  $T_f$  in the case of the relaxation to the  Gibbs equilibrium state is deduced from the diffusion coefficients~\cite{hamd1,hamd2}:
 \begin{eqnarray}
  D_{xx}&=&\frac{\hbar \lambda }{4\pi m } \int_{-\pi}^{\pi} \frac{\coth\left(\frac{\hbar \omega(q)}{2k_{B}T_f}\right)}{\omega(q)} dq,\\
   D_{pp}&=&\frac{\hbar m \lambda }{4\pi } \int_{-\pi}^{\pi} \coth\left(\frac{\hbar \omega(q)}{2k_{B}T_f}\right)\omega(q)dq ,\\
    D_{\rm{ex}}&=&\frac{\hbar \lambda }{4\pi m } \int_{-\pi}^{\pi} \frac{\coth\left(\frac{\hbar \omega(q)}{2k_{B}T_f}\right)}{\omega(q)} \cos(q) dq,
 \end{eqnarray}
where $k_B$ is the Boltzmann constant and 
\begin{equation}
 \omega(q)=\sqrt{\omega_0^2+\frac{4\xi}{m}\sin^2(q/2)}.
\end{equation}
At high temperatures, the above  integrals simplify to:
\begin{eqnarray}
  D_{xx}&=&\frac{\lambda k_B T_f}{m\omega_0\sqrt{\omega_0^2+\frac{4\xi}{m}}},\\
   D_{pp}&=& m\lambda k_B T_f,\\
    D_{\rm{ex}}&=& \frac{\lambda k_B T_f}{\xi \omega_0} \left(\frac{\omega_0^2+\frac{2\xi}{m}-\omega_0\sqrt{\omega_0^2+\frac{4\xi}{m}}}{\sqrt{\omega_0^2+\frac{4\xi}{m}}}\right).
   \end{eqnarray}
   Taking into account these coefficients, it follows from equation~(\ref{sink}) that:
   \begin{equation}
    s(T_f)=\frac{2\lambda k_B T_f}{a}.
   \end{equation}
The heat transport equation becomes:
\begin{eqnarray}
 \frac{\partial u(x,t)}{\partial t}=\left(\frac{ a^2 \xi}{2\lambda m}\right)\frac{\partial^2 u(x,t)}{\partial x^2}-2\lambda\left( u(x,t)- \frac{k_B T_f}{a}\right).
\end{eqnarray}
The latter result has a simple statistical explanation: At high temperature, by the equipartition theorem, the  equilibrium mean energy of the lattice is:
\begin{equation}
 \overline U=Nk_BT_f.
\end{equation}
Hence the equilibrium energy density reads:
\begin{equation}
 u_{\rm eq}=\frac{Nk_BT_f}{Na}=\frac{k_B T_f}{a}= \frac{s(T_f)}{2\lambda},
\end{equation}
which will be reached in the continuum limit once $\frac{\partial u(x,t)}{\partial t}=\frac{\partial^2 u(x,t)}{\partial x^2}=0$.
Furthermore, at such temperatures, the heat capacity density behaves classically and is thus nearly constant:
\begin{equation}
 C(T)=\frac{k_B}{a},
\end{equation}
whereas the thermal conductivity takes the form:
\begin{equation}
 \kappa=\frac{ a k_B \xi}{2\lambda m}=\frac{  k_B \xi}{2\lambda \rho},
\end{equation}
where $\rho$ is the mass density of the lattice.
The heat transport equation simplifies further to
\begin{eqnarray}
 \frac{\partial T(x,t)}{\partial t}=\sigma \frac{\partial^2 T(x,t)}{\partial x^2}-2\lambda\left( T(x,t)-  T_f\right),\label{fin}
\end{eqnarray}
where $\sigma$ is the thermal diffusivity: 
\begin{equation}
 \sigma=\frac{\kappa a}{k_B}=\frac{a^2\xi}{2\lambda m}.
 \end{equation}
  Obviously, the diffusivity is proportional to the coupling constant $\xi$ and is inversely proportional to the atoms mass. The second term on the right-hand side of equation~(\ref{fin}) expresses the fact that the rate of change of temperature between the chain and the reservoir is proportional to the difference between the temperatures of the two systems, which is Newton's classical law of heating. 

\section{Conclusion}
To sum up, we have derived the  heat transport equation in dielectrics and semiconductors  starting from a Markovian master equation for the one-dimensional lattice.
We first derived the equations of motion for the discrete lattice and discussed the contribution of the energy currents at the microscopic level. We found that the range over which these currents influence energy transport extends beyond the lattice constant, a fact that depends on both the phonons decay time and the speed of propagation.  The effect of the environment is shown to be equivalent to a  source density that depends on the diffusion coefficients and implicitly on the decay time.  We evaluated the rate of change of the energy density in the continuum limit and showed that in this limit, which is associated with long wavelength phonons, the speed of propagation of heat is constant, which turns out to be the speed of propagation of sound in the lattice. The thermal conductivity takes a form similar to that of gases, and depends on the temperature through the heat capacity density of the lattice. At high temperatures, the latter is constant, and by calculating the exact form of the diffusion coefficients, we recover the  classical heat equation.

It is quite interesting to extend the investigation to the case of strong coupling between the lattice and the environment, where quantum fluctuations play an important role. In this regime,  the dynamics is most likely to be non-Markovian at low temperatures, and the exact treatment of the non-Markovian dynamics  is still an open problem that is not fully solved even for systems with few degrees of freedom. One can invoke for example the path integral technique which is cumbersome even for one oscillator. One of the main features of the non-Markovian dynamics is the time dependence of the damping coefficients and eventually of the diffusion coefficients. The energy currents in this case may display very complicated behavior due to memory effects which certainly affects the thermal conductivity. However, at high temperatures, under the assumption that memory effects and quantum fluctuations become negligible,  one expects that the dynamics reproduces the classical thermal transport equations as is the case for weak coupling which is the main aim of this work. 
\section*{Appendix}
For nonzero values of the damping coefficient $\gamma$, the diffusion coefficients at arbitrary temperature read:
 \begin{eqnarray}
  D_{xx}&=&\frac{\hbar }{4\pi m } \int_{-\pi}^{\pi} \frac{\coth\left(\frac{\hbar \omega(q)}{2k_{B}T_f}\right)}{\omega(q)} [ \lambda+2\gamma \cos(q)]dq,\\
   D_{pp}&=&\frac{\hbar m  }{4\pi } \int_{-\pi}^{\pi} \coth\left(\frac{\hbar \omega(q)}{2k_{B}T_f}\right)\omega(q)  [ \lambda+2\gamma \cos(q)]dq ,\\
    D_{\rm{ex}}&=&\frac{\hbar }{4\pi m } \int_{-\pi}^{\pi} \frac{\coth\left(\frac{\hbar \omega(q)}{2k_{B}T_f}\right)}{\omega(q)} \cos(q)  [ \lambda+2\gamma \cos(q)] dq.
 \end{eqnarray}
In the limit of high temperature, we find:
\begin{eqnarray}
  D_{xx}&=&\frac{\lambda k_B T_f}{m\omega_0\sqrt{\omega_0^2+\frac{4\xi}{m}}}+\frac{\gamma k_B T_f}{\xi \omega_0} \left(\frac{\omega_0^2+\frac{2\xi}{m}-\omega_0\sqrt{\omega_0^2+\frac{4\xi}{m}}}{\sqrt{\omega_0^2+\frac{4\xi}{m}}}\right) ,\\
   D_{pp}&=& m\lambda k_B T_f,\\
    D_{\rm{ex}}&=& \frac{\lambda k_B T_f}{\xi \omega_0} \left(\frac{\omega_0^2+\frac{2\xi}{m}-\omega_0\sqrt{\omega_0^2+\frac{4\xi}{m}}}{\sqrt{\omega_0^2+\frac{4\xi}{m}}}\right)\nonumber \\
    &+& \frac{(2\gamma k_B T_f)/(m\omega_0^2)}{1+\frac{2\xi }{2\xi+m\omega_0^2}+\sqrt{1+\frac{4\xi}{m \omega_0^2}}}.
   \end{eqnarray}
   By direct substitution into equation~(\ref{sink}) we find once again that:
   \begin{equation}
    s(T_f)=\frac{2\lambda k_B T_f}{a}.
   \end{equation}

   \end{document}